\documentclass[twocolumn]{aastex631}

\usepackage{hyperref}
\begin{document}

\title{SPOCK 2.0:\\ Updates to the FeatureClassifier in the Stability of Planetary Orbital Configurations Klassifier}

\author[0009-0000-4343-4253]{Elio Thadhani}
\affiliation{Department of Physics, Harvey Mudd College, Claremont, CA 91711, USA}

\author[0009-0008-5374-5803]{Yanming Ba}
\affiliation{Department of Physics, Harvey Mudd College, Claremont, CA 91711, USA}

\author[0000-0003-1927-731X]{Hanno Rein}
\affiliation{Department of Physical and Environmental Sciences, Univeresity of Toronto at Scarborough, Toronto, Ontario, M1C 1A4, Canada}
\affiliation{Department of Astronomy \& Astrophysics, Univeresity of Toronto, Toronto, Ontario, M5S 3H4, Canada}

\author[0000-0002-9908-8705]{Daniel Tamayo}
\altaffiliation{Corresponding author: \href{mailto:dtamayo@hmc.edu}{dtamayo@hmc.edu}}
\affiliation{Department of Physics, Harvey Mudd College, Claremont, CA 91711, USA}

\begin{abstract}
The Stability of Planetary Orbital Configurations Klassifier (SPOCK) package collects machine learning models for predicting the stability and collisional evolution of compact planetary systems. 
In this paper we explore improvements to SPOCK's binary stability classifier (FeatureClassifier), which predicts orbital stability by collecting data over a short N-body integration of a system.
We find that by using a system-specific timescale (rather than a fixed $10^4$ orbits) for the integration, and by using this timescale as an additional feature, we modestly improve the model's AUC metric from 0.943 to 0.950 (AUC=1 for a perfect model). We additionally discovered that $\approx 10\%$ of N-body integrations in SPOCK's original training dataset were duplicated by accident, and that $<1\%$ were misclassified as stable when they in fact led to ejections.
We provide a cleaned dataset of 100,000+ unique integrations, release a newly trained stability classification model, and make minor updates to the API.

\end{abstract}

\section{Introduction} \label{sec:intro}
Determining orbital stability over planetary systems' typical lifetimes of several Gyr through direct numerical integration is computationally expensive. As an alternative, the FeatureClassifier in the Planetary Orbital Configurations Klassifier (SPOCK) package \citep{Tamayo2020} uses machine learning to rapidly predict the stability of compact orbital configurations. The FeatureClassifier extracts ten dynamically relevant features \citep{spockI} over the span of short $10^4$-orbit N-body integrations, and passes them to a trained model which predicts stability over $10^9$ orbits\footnote{All references to the number of orbits refer to a system's smallest orbital period.} based on these features.



In this paper, we relax the rigid $10^4$-orbit integration time applied to all systems and explore implementing a system-specific timescale, both as the integration time and as a model feature.


\section{Updated Dataset and Metrics} \label{sec:update}

\cite{Tamayo2020} used all of the holdout testing systems when reporting accuracy metrics; however, some of these systems go unstable during the short integration phase, obviating the need for a prediction from SPOCK.
We therefore redefine our metrics to exclude systems that go unstable during the short integration phase; which slightly reduces the model AUC from 0.9527 to 0.9490 (an AUC of 1 would be a perfect model). 

\subsection{Cleaned Dataset}

SPOCK was trained on a dataset of $\sim 10^5$ resonant and near-resonant systems. A holdout subset of this ``resonant" dataset, as well as a ``random" dataset of $25,000$ randomly generated configurations was used for testing.
Since SPOCK's release, we have found that $\approx 10\%$ (10,984) of systems in the resonant dataset were duplicated (none in the random dataset). 
We additionally find that the code missed a small subset of ejections, and $<1\%$ of systems that lost a planet were incorrectly labeled as stable (65 in resonant and 59 in random datasets). 
We have uploaded a corrected dataset to Zenodo \citep{cleanDataset} with a total of 102,497 and 24,941 unique systems in the resonant and random datasets respectively . 


We find that a model trained and tested on the original dataset has an AUC of 0.9490, whereas a model trained and tested on the cleaned dataset has an AUC of 0.9426. 
This slight performance drop is expected given that we removed systems that the model had seen during training.

Additionally, since the current version of REBOUND can no longer read the simulation archive files released in the original paper, we have repackaged the initial conditions for each dataset into a single binary file of simulation archives readable as of REBOUND version 4.4.6 \citep{Rein17}. 
We note that, due to finite floating point precision and intervening changes in the integration algorithms, running these initial conditions in the latest version of REBOUND does not yield identical trajectories and instability times as in \cite{Tamayo2020}.
However, the new trajectories and instability times represent equally valid realizations of the chaotic dynamics, i.e., a collection of trajectories perturbed at the level of the machine precision will yield a distribution of instability times with a well-defined mean and spread, both of which are set by chaotic dynamics \citep{Hussain20}. 
In the original integrations of \cite{Tamayo2020}, one such ``shadow trajectory" was run for each initial condition, and we have confirmed that, e.g., running the short integrations with one trajectory while predicting the instability times of the shadow trajectories does not degrade the model performance.

\section{SPOCK 2.0}

\subsection{REBOUND changes: MEGNO}

One of the ten features in our model is the MEGNO chaos indicator \citep{Cincotta03}.
Due to the rapid divergence of chaotic trajectories, the MEGNO for some systems in our original training dataset overflowed to produce \texttt{NaN} values. 
New versions of REBOUND fix this issue by periodically renormalizing the phase-space distance between the original and shadow trajectory. 
We find that the original SPOCK model and a model trained with the new version of MEGNO calculations have similar performance (testing AUC of 0.9426 and 0.9430 respectively). 
We believe this is because MEGNO values overflowed only in strongly chaotic systems, which were the easiest cases to predict.

\subsection{Short integration timescale}

In the original FeatureClassifier all initial conditions are run for $10^4$ orbits for the short integration. However, the dynamical timescales for low-mass, widely separated planets are longer than for high-mass, closely packed planets, creating an inconsistency in system behavior during our evaluation window.
We therefore test whether tailoring the short integration time to individual systems improves performance. 
\cite{analytical} argue that slow ``secular" variations in the orbital eccentricities can modulate the extents of mean motion resonances (MMRs) and the extent to which they overlap with one another, driving chaos. 
An N-planet system will have N secular frequencies \citep[e.g.,][]{Murray99}, so we try running each system out to the timescale of its fastest secular frequency\footnote{This choice balances capturing some of the secular dynamics against keeping the N-body integrations short to decrease evaluation time. Additionally, the longest secular timescale tends to infinity in the closely packed limit, where the slowest mode becomes equivalent to the ``center-of-mass eccentricity vector" \citep{Yang24}, which is conserved in this limit \citep{Goldreich81}.}.

To do this we use the analytic approximations by \cite{Yang24} for three-planet systems (for higher multiplicity systems, we calculate this timescale for all adjacent trios and take the minimum secular timescale). 
The fastest secular timescale $T_{sec}$ is approximately given by 
\begin{equation}
    T_{sec} = 2\pi \Bigg(\frac{m_\star}{m_1+m_2+m_3}\Bigg)\frac{P_3}{\xi}
\end{equation} 
where $P_3$ is the orbital period of the third (outermost) planet, $m_i$ is the mass of the $i$th planet, and $m_\star$ is the mass of the star.
The quantity $\xi$ accounts for the orbital separations and distribution of mass between the innermost and outermost planet,
\begin{equation}
\xi = \frac{m_1}{m_1+m_3}\frac{1}{e_{c,12}^2} + \frac{m_3}{m_1+m_3}\frac{1}{e_{c,23}^2}
\end{equation}
where the crossing eccentricities are defined in terms of the semimajor axes $a_i$ as
\begin{equation}
    e_{c,12} = \left(\frac{a_1}{a_2}\right)^{-\frac{1}{4}} \left(\frac{a_2}{a_3}\right)^{\frac{5}{8}}\left(1-\frac{a_1}{a_2}\right) \approx \frac{a_2-a_1}{a_2} \label{eq:e12}
\end{equation}
\begin{equation}
    e_{c,23} = \left(\frac{a_1}{a_2}\right)^{\frac{1}{8}}\left(\frac{a_2}{a_3}\right)^{-\frac{1}{2}} \left(1-\frac{a_2}{a_3}\right) \approx \frac{a_3-a_2}{a_3},\label{eq:e23}
\end{equation}
 with the approximate equalities holding in the closely spaced limit.

We also experimented with numerical solutions to this secular timescale using the \texttt{celmech} code \citep{celmech}, but found no significant difference in performance, so we adopt the above analytic expressions that can be quickly evaluated.


We collect feature data over short integrations to a system's fastest secular timescale ($T_{sec}$). 
We note that the short N-body integration determines stability up to time $T_{sec}$, so we only train and test on systems that survive beyond $T_{sec}$ (Sec.\:\ref{sec:update}). 
Additionally, we try adding $T_{sec}$ as a model feature. 
The results are summarized in table \ref{tab:modelCompare}. 
\begin{table}[h!]

    \centering
    \begin{tabular}{|c|c|c|}
    \hline
        Model comparison & AUC & FPR \\
        \hline
        Run to 1e4 with old features & 0.9430 & 0.1605\\
    \hline
        Run to $T_{sec}$ with old features & 0.9417 & 0.1660\\
        \hline
    Run to 1e4 with $T_{sec}$ as feature & 0.9442 & 0.1595\\
    \hline
    Run to $T_{sec}$ with $T_{sec}$ as feature & 0.9502 & 0.1431\\
    \hline
    \end{tabular}
    \caption{Summary of results comparing running short N-body integrations to a fixed $10^4$ orbits vs. to $T_{sec}$ (varying by system). We also compare including $T_{sec}$ as an additional feature to the model.}
    \label{tab:modelCompare}
\end{table}

We find that integrating to $T_{sec}$ and including this value as an additional feature increases the AUC by a small but significant amount ($\gtrsim 10\%$ of the gap between the previous SPOCK model and AUC=1).

Note, the median $T_{sec}$ of our training/testing dataset is $1.15\times 10^4\times P_{min}$, with 98.4 \% of the systems in our training set falling between $10^3$ and $10^5$ orbits.
This means our updated model's median evaluation time is $\approx 15\%$ slower than the previous model, but now varies by system.

\subsection{Code Release}


We release SPOCK 2.0 with a retrained FeatureClassifier model, using our cleaned dataset, integrating to the input system's fastest secular time scale, and using that timescale as a model feature - modestly improving model performance.

\begin{acknowledgments}
\section{Acknowledgments}
We thank Eritas Yang for her guidance implementing the secular timescale calculations and Connor Neely for his assistance updating the model.
\end{acknowledgments}



\nocite{Chen:2016:XST:2939672.2939785,harris2020array,reback2020pandas,Hadden18,analytical}
\bibliography{spock}{}
\bibliographystyle{aasjournal}



\end{document}